\documentclass[twocolumn,12pt,preprint]{aastex61}
\usepackage{amsmath}
\usepackage{ulem}
\usepackage{xcolor}

\newcommand{\cri}{_{\rm cr}}

\newcommand{\Be}{_{\rm Be}}

\newcommand{\kuzndll}{\texttt{libGS\_Std\_HomSrc\_CEH}}
\newcommand{\fldll}{\texttt{libGS\_Std\_HomSrc\_C}}

\shorttitle{Ultimate Fast GS Codes}
\shortauthors{Kuznetsov \& Fleishman}

\begin{document}
\title{Ultimate Fast Gyrosynchrotron Codes}

\author[0000-0001-8644-8372]{Alexey A. Kuznetsov}
\affiliation{Institute of Solar-Terrestrial Physics, Irkutsk, 664033, Russia}

\author[0000-0001-5557-2100]{Gregory D. Fleishman}
\affiliation{Physics Department, Center for Solar-Terrestrial Research, New Jersey Institute of Technology
Newark, NJ, 07102-1982}
\affil{Ioffe Institute, Polytekhnicheskaya, 26, St. Petersburg, 194021, Russia}

\begin{abstract}
The past decade has seen a dramatic increase of practical applications of the microwave gyrosynchrotron emission for plasma diagnostics and three-dimensional modeling of solar flares and other astrophysical objects. This break-through turned out to become possible due to apparently minor, technical development of Fast Gyrosynchrotron Codes, which enormously reduced the computation time needed to calculate a single spectrum, while preserving accuracy of the computation. However, the available fast codes are limited in that they could only be used for a factorized distribution over the energy and pitch-angle, while the distributions of electrons over energy or pitch-angle are limited to  a number of pre-defined analytical functions. In realistic simulations, these assumptions do not hold; thus, the codes free from the mentioned limitations are called for.   To remedy this situation, we extended our fast codes to work with an arbitrary input distribution function of radiating electrons. We accomplished this by implementing fast codes for a distribution function described by an arbitrary numerically-defined array. In addition, we removed several other limitations of the available fast codes and improved treatment of the free-free component. The Ultimate Fast Codes presented here allow for an arbitrary combination of the analytically and numerically defined distributions, which offers the most flexible use of the fast codes. We illustrate the code 
with a few simple examples.
\end{abstract}

\keywords{radiation mechanisms: non-thermal---methods:
numerical---Sun: radio radiation---Sun: flares---stars: flares---radio continuum: planetary systems}


















\section{Introduction}\label{Sec_intro}


Radiation produced by moderately relativistic electrons gyrating in the ambient magnetic field, commonly called gyrosynchrotron (GS) emission, makes a dominant contribution to the microwave emission of solar and stellar flares and is important in other astrophysical objects. This emission process is well understood theoretically; the exact formulae for the emissivities and absorption coefficients
\citep{Eidman_1958, Eidman_1959, Melrose_1968, Ramaty_1969} are broadly applicable for arbitrary conditions provided that the magnetic field changes in space only smoothly \citep[otherwise, if the magnetic field contains sharp changes, a different kind of emission, \textit{diffusive synchrotron radiation}, is produced,][]{Fl_2006a,Li_Fl_2009} and no quantum effect is important. The problem with those exact formulae is that they are extremely slow computationally, especially, when high harmonics of the gyrofrequency are involved. The computation speed {often} matters, but becomes critical in two practically important cases. One of them is three-dimensional (3D) modeling when emission in many (up to a few hundred thousand) elementary model volumes (voxels) has to be computed \citep{Nita_etal_2015}. Another one is the model spectral fitting, when multiple trial spectra are computed to identify a theoretical spectrum that best matches the observed one \citep{2020Sci...367..278F}.
The solution to this problem was obtained by development of Fast GS Codes by \citet{Fl_Kuzn_2010} following \cite{Petrosian_1981} and \cite{Klein_1987}.


The development of the Fast GS Codes enabled a break-through in massive 3D modeling of the microwave emission from solar \citep[e.g.,][]{Fl_etal_2016,Fl_etal_2017,2017A&A...604A.116G,2018ApJ...852...32K,2018ApJ...859...17F,2019AdSpR..63.1453G,2020ApJ...890...75M,2021ApJ...913...97F} and stellar \citep{2019MNRAS.483..917W, 2020MNRAS.496.2715W} flares. Most of these studies employed a dedicated modeling tool, GX Simulator \citep{Nita_etal_2015, Nita_etal_2018}, designed specifically to simulate emissions from solar flares and active regions \citep{2021ApJ...909...89F}. A limitation of the Fast GS Codes \citep{Fl_Kuzn_2010} and, thus, of the GX Simulator, is in the allowable distribution function of the nonthermal electrons. The codes assumed a factorized distribution over the energy and the pitch-angle, where the user might select from several predefined analytical options. Although this was sufficient for many practical applications, such an approach cannot account for the whole variety of cases relevant to the physics involved in the particle acceleration and transport in solar flares.


New microwave imaging spectroscopy data \citep[e.g.,][]{Gary_etal_2018} and new numerical simulations \citep[e.g.,][]{PhysRevLett.126.135101} call for a more general treatment free from the mentioned limitation. Specifically, the ability to compute the GS emission produced by a nonthermal electron distribution represented by an arbitrary numerical array is needed. In the general case such array would describe a non-factorized anisotropic particle distribution with an arbitrary energy spectrum.


There are ample evidence that real distributions of the nonthermal particles cannot be properly described in a factorized form. This is clear from the fact that a particle isotropisation time depends on energy for most of the possible scattering processes---both collisional (due to Columb collisions) and collisionless (e.g., due to waves). Thus, even if the electron distribution is somehow created in a factorized form, it will immediately become non-factorized due to transport effects.


Numerical solutions of transport equations naturally come in the form of arrays. Therefore, extension of the available fast codes to a more general case of the array-defined distributions is demanded by the current state-of-the-art of science.

This paper describes such a new extension of the fast codes that permits electron distributions to be described by an arbitrary numerical array. In addition to the ability to work with an arbitrary array-defined distributions, the new codes also include several new useful features. 
The current version of the fast codes includes all relevant physical processes, which we were able to foresee; thus, we call them Ultimate Fast GS Codes.

\section{Features}
Here, we briefly describe the key features implemented in the previous releases of the codes, and introduce the new capabilities of the Ultimate Fast GS Codes.

\subsection{Previously implemented features}
The key feature of the fast GS codes is the approximate continuous algorithm to compute the gyrosynchrotron emissivities and absorption coefficients for the ordinary and extraordinary electromagnetic modes. This algorithm is described in detail by \citet{Fl_Kuzn_2010}; it is based on replacing the summation over cyclotron harmonics $s$ by integration over them. After that, the expressions for the emissivity and absorption coefficient are reduced to two-dimensional integrals over the particle energy and pitch-angle, and  integration over pitch-angle is performed using the Laplace integration method \citep{Petrosian_1981}. High accuracy of this integration is ensured by a close proximity of the integrand to the Gaussian function, for which the Laplace method gives the exact result. The code provides two slightly different implementations of the continuous algorithm, optimized either for speed or for accuracy. In general, the continuous approximation improves the computation speed by orders of magnitude (in comparison to the exact formulae), while providing a reasonably high accuracy at high frequencies (at $f\gg f_{\mathrm{B}}$, {where $f_{\mathrm{B}}=eB/(2\pi m c)$ is the electron cyclotron (gyro) frequency; $e$ and $m$ are the charge and mass of the electron, respectively, $c$ is the speed of light, and $B$ is the magnetic field strength}). However, by construction, this algorithm does not reproduce the harmonic structure of the emission at low frequencies.

To remedy the above limitation (if necessary), the code provides the capability to compute the gyrosynchrotron emission parameters at low frequencies using the exact formalism by \citet{Melrose_1968}; the boundary frequency, below which the exact computation is applied, is specified by the user (in units of the cyclotron frequency). In turn, the exact gyrosynchrotron code can use either the exact values of Bessel functions, or the approximation by \citet{1971AuJPh..24...43W}, which provides either a higher accuracy or a higher computation speed, respectively. The exact gyrosynchrotron formulae (at a single frequency) can also be used to obtain correction factors for the fast continuous algorithm, which further improves the accuracy of that algorithm.

{Although both the exact gyrosynchrotron formulae and the continuous approximation can be applied to arbitrary electron distribution functions, the first implementation of the fast code, for simplicity, used only analytical distribution functions specified in the factorized form: $f(E, \mu)=u(E)g(\mu)$, where $E$ is the electron kinetic energy, $\mu=\cos\alpha$, and $\alpha$ is the electron pitch-angle. The code  contains} a number of built-in model distributions over energy (such as thermal, {kappa,} power law, double power law, etc.) and over pitch-angle (loss-cone, beam-like, etc.) which can be arbitrarily combined to create a diverse  (but, as said above, still limited) set of two-dimensional electron distributions.

The code accounts for the contribution of the free-free (FF) emission mechanism, i.e., the emissivity ($j_{\sigma}$) and absorption coefficient ($\varkappa_{\sigma}$) for the emission mode $\sigma$ are computed as $j_{\sigma}=j_{\sigma}^{\mathrm{GS}}+j_{\sigma}^{\mathrm{FF}}$ and $\varkappa_{\sigma}=\varkappa_{\sigma}^{\mathrm{GS}}+\varkappa_{\sigma}^{\mathrm{FF}}$. The free-free emission mechanism includes  contributions of the electron-ion and electron-neutral collisions (the latter may dominate in the chromosphere).

The code computes the emission intensity $I$ from an inhomogeneous source by numerical integration of the one-dimensional (thus, no radiation scattering is included) radiation transfer equations along the line of sight, separately for each frequency.\footnote{This capability was not yet available in the first code release by \citet{Fl_Kuzn_2010}, while implemented  later, as described, e.g., by \citet{Nita_etal_2015}.} For the most part of a line of sight $z$, away from regions with quasi-transverse magnetic field, the electromagnetic emission modes propagate independently from each other and with different group velocities \citep{Fl_etal_2002}, so that the radiation transfer equations for these modes have the form
\begin{equation}
\frac{\mathrm{d}I_{\mathrm{R,L}}(z)}{\mathrm{d}z}=j_{\mathrm{R,L}}(z)-\varkappa_{\mathrm{R,L}}(z)I_{\mathrm{R,L}}(z),
\end{equation}
where the indices R and L refer to the right and left elliptically polarized components, respectively; at each location, these components correspond to either ordinary or extraordinary electromagnetic mode, depending on the direction of the magnetic field vector relatively to the line of sight. If the emission crosses a layer of transverse magnetic field (i.e., where the projection of the magnetic field vector on the line of sight changes sign), partial mode conversion can occur, which is described by the relations
\begin{eqnarray}
I_{\mathrm{R}}^{\mathrm{out}} & = & Q_{\mathrm{T}}I_{\mathrm{R}}^{\mathrm{in}}+(1-Q_{\mathrm{T}})I_{\mathrm{L}}^{\mathrm{in}},\nonumber\\
I_{\mathrm{L}}^{\mathrm{out}} & = & Q_{\mathrm{T}}I_{\mathrm{L}}^{\mathrm{in}}+(1-Q_{\mathrm{T}})I_{\mathrm{R}}^{\mathrm{in}},
\end{eqnarray}
where \citep{1960ApJ...131..664C, 1964SvA.....7..485Z}
\begin{equation}
Q_{\mathrm{T}}=2^{-f_{\mathrm{T}}^4/f^4},\qquad
f_{\mathrm{T}}^4=\frac{\pi^2}{4c\ln 2}\frac{f_{\mathrm{p}}^2f_{\mathrm{B}}^3}{|\mathrm{d}\theta/\mathrm{d}z|},
\end{equation}
{$f_{\mathrm{p}}=e\sqrt{n_{\mathrm{e}}/(\pi m)}$ is the electron plasma frequency, $n_{\mathrm{e}}=n_0 +n_{\mathrm{b}}$ is the total number density of the thermal ($n_0$) and nonthernal ($n_{\mathrm{b}}$) free electrons}, and the gradient of the viewing angle along the line of sight $\mathrm{d}\theta/\mathrm{d}z$ is computed numerically. The depolarization frequency $f_{\mathrm{T}}$ is defined in such a way that for $f=f_{\mathrm{T}}$ we obtain $Q_{\mathrm{T}}=0.5$ and the emission exiting the transverse magnetic field layer becomes unpolarized regardless of the emission polarization before that layer. In addition to the above formalism (which we call ``exact'' coupling), the code implements, for {the} testing, the ``weak'' ($Q_{\mathrm{T}}\equiv 0$) and ``strong'' ($Q_{\mathrm{T}}\equiv 1$) coupling modes (equivalent to $f\ll f_{\mathrm{T}}$ and $f\gg f_{\mathrm{T}}$, respectively).

\subsection{New capabilities}
The key enhancement of the Fast GS Codes reported here is the possibility to use arbitrary electron distribution functions. These functions are specified {via} two-dimensional arrays $f_{ij}=f(E_i, \mu_j)$, along with the corresponding energy and pitch-angle grids. {The values of the distribution function $f(E, \mu)$ between the grid nodes, as well as its partial derivatives over the energy and pitch-angle, are computed using two-dimensional interpolation (see Appendix \ref{Interpolation}).} {This approach allows using non-factorized electron distribution functions (and the distributions that cannot be reduced to linear combinations of factorized ones), including, e.g., distributions with energy-dependent anisotropy of the electrons or, complementary, energy distributions that depend on the pitch-angle; see examples in Fig.\,\ref{FigDFs} below.} The gyrosynchrotron emission parameters are computed using exactly the same approach as in the previous releases (i.e., either the fast continuous algorithm by \citealt{Fl_Kuzn_2010}, or the exact formalism by \citealt{Melrose_1968}, or a combination of them); The computation speed and the resulting accuracy have been found to be comparable to those for similar analytical distribution functions.

The code retains the full functionality of the previous versions. In particular, it can use either array-defined or analytical distribution functions, or a combination of them. In the latter case, the gyrosynchrotron emissivities and absorption coefficients are computed separately for the array-defined and analytical components of the electron distribution, and then added together: $j_{\sigma}^{\mathrm{GS}}=j_{\sigma}^{\mathrm{GS,\, ana}}+j_{\sigma}^{\mathrm{GS,\, arr}}$, $\varkappa_{\sigma}^{\mathrm{GS}}=\varkappa_{\sigma}^{\mathrm{GS,\, ana}}+\varkappa_{\sigma}^{\mathrm{GS,\, arr}}$ (the computation speed decreases accordingly).

Other improvements include a more accurate treatment of the free-free emission component, according to the new formalism by \citet{2021ApJ...914...52F}; in particular, there is a possibility to choose the element abundances typical of either the solar corona or the chromosphere. The code provides the capability to use either a user-defined array of frequencies where the emission is computed, or an automatically computed (logarithmically spaced) array. The list of built-in analytical electron distributions has been expanded. Finally, some bugs have been fixed and several numerical algorithms have been optimized.


There are several practical considerations, which are helpful to take into account while using the ultimate codes. For the numerically-defined distributions there is not too much room for improving accuracy of the integration (primarily, over energy) because the number and values of integration nodes are dictated by the array itself. This means that the user has to take care of the accuracy while generating the array from their specific numerical model. In particular, for a power-law like distributions, it is preferable that the energy nodes are logarithmically spaced. In this case, the trapezoidal integration in the log-space will give the best results (exact in case of a true power-law), which is the default integration option of the codes.



\section{Code implementation}\label{sec_Appl}

The key {new capability---the ability of using an arbitrary non-factorized numerical distribution function as described above---}has been implemented in two entirely independent codes that employ either FORTRAN or C++ to cross-validate them  similarly to \citet{Fl_Kuzn_2010}, but we release only one of them, C++, to avoid any confusion of the perspective users. The Ultimate Fast GS Codes have been implemented as executable libraries (dynamic link libraries for Windows and shared libraries for Linux or MacOS) callable from IDL or Python. The C++ source code together with the compiled libraries, complete description, and examples of the use cases are available on GitHub\footnote{\url{https://github.com/kuznetsov-radio/gyrosynchrotron}}; version 1.0.0 is archived in Zenodo \citep{Kuznetsov2021}. The code can compute the emission parameters either for a single line of sight, or for a set of lines of sight simultaneously (using the capabilities of multi-processor systems).

The new codes have been integrated into the 3D modeling and simulation tool GX Simulator \citep{Nita_etal_2015, Nita_etal_2018}; the new capabilities of this tool and possible applications will be presented elsewhere. The Ultimate Fast GS Codes can also be used as a standalone application provided that the user properly supplies the necessary input parameters.


\section{Application: energy-dependent loss-cone distribution}
To illustrate the capabilities of the new codes, we have applied them to a particular type of non-factorizable electron distribution. We assume that energetic electrons are injected at large pitch angles and have initially a highly-anisotropic ``pancake''-like distribution with a sharp peak at $\alpha=90^{\circ}$. Considering the pitch-angle scattering only, we can write the kinetic equation for the distribution function as \citep{Fl_Topt_2013_CED} 
\begin{figure*}
\includegraphics{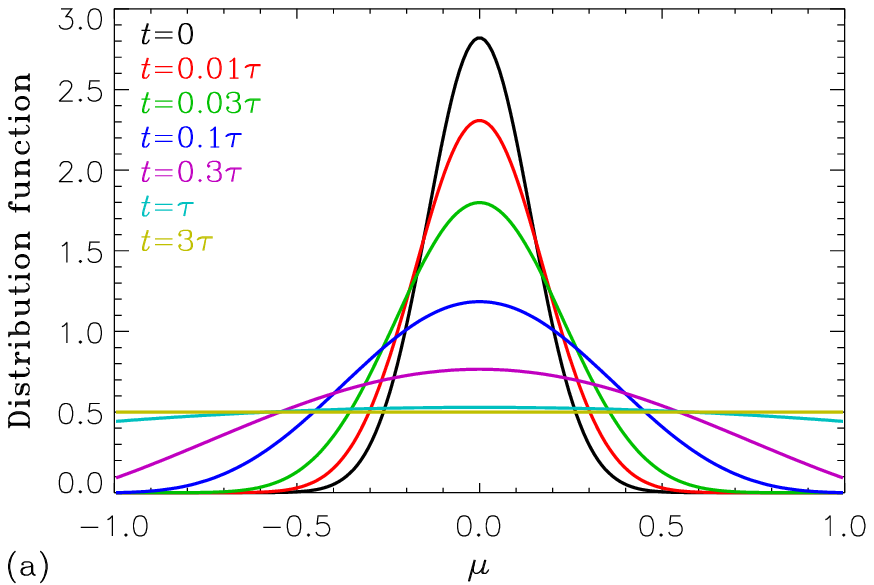}%
\includegraphics{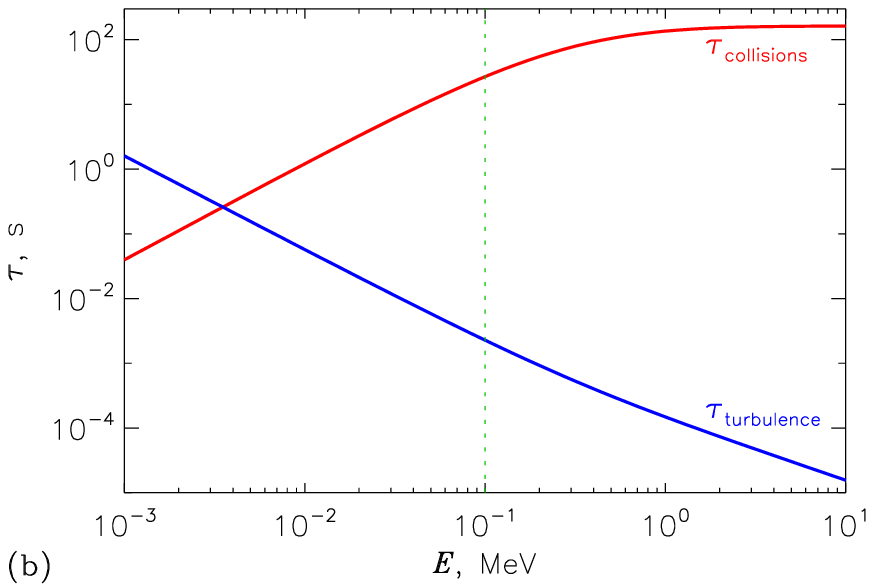}
\caption{(a) Time evolution of the electron distribution function. (b) Energy dependence of the particle scattering time for two scattering models: scattering due to collisions (in a plasma with the electron density of $10^{10}$ $\textrm{cm}^{-3}$ and temperature of $10^7$ K) and scattering on turbulence \protect\citep[with the parameters reported by][]{2018A&A...610A...6M}; the vertical dotted line marks the low-energy cutoff of the nonthermal electron distribution used in this work.}
\label{FigEvolutionTau}
\end{figure*}
\begin{equation}\label{mu_scattering}
\frac{\partial f}{\partial t}=\frac{\partial}{\partial\mu}\left(D_{\mu\mu}\frac{\partial f}{\partial\mu}\right),
\end{equation}
where {the angular diffusion coefficient $D_{\mu\mu}$ has the form}
\begin{equation}
D_{\mu\mu}=\frac{\nu}{2}(1-\mu^2)
\end{equation}
and $\nu=1/\tau$ is the scattering rate, which is reciprocal to the isotropization time $\tau$. The distribution function changes with time as shown in Figure \ref{FigEvolutionTau}a, and becomes nearly isotropic at $t\ge\tau$. Since the scattering rate $\nu=\nu(E)$ is energy-dependent, the resulting pitch-angle distribution will be energy-dependent, too. For different conditions, this process can produce qualitatively different shapes of the electron distribution which are investigated below.

\subsection{Collisional scattering}
Scattering due to collisions with the ambient particles has the following characteristics \citep[e.g.,][]{1965RvPP....1..105T, Fl_Topt_2013_CED}:
\begin{equation}\label{tau_c}
\tau_{\mathrm{c}}=\frac{m^2v^3}{4\pi n_0e^4\ln\Lambda},\qquad
\ln\Lambda=\ln\frac{8\times 10^6 T}{\sqrt{n_0}},
\end{equation}
{where $v$ is the electron velocity, and $n_0$ and $T$ are the plasma density and temperature, respectively;} i.e., the \textit{low-energy} electrons are scattered more efficiently. An example for the energy dependence of the collisional scattering time $\tau_{\mathrm{c}}(E)$ (for $n_0=10^{10}$ $\textrm{cm}^{-3}$ and $T=10^7$ K) is shown in Figure \ref{FigEvolutionTau}b.

\begin{figure*}
\includegraphics{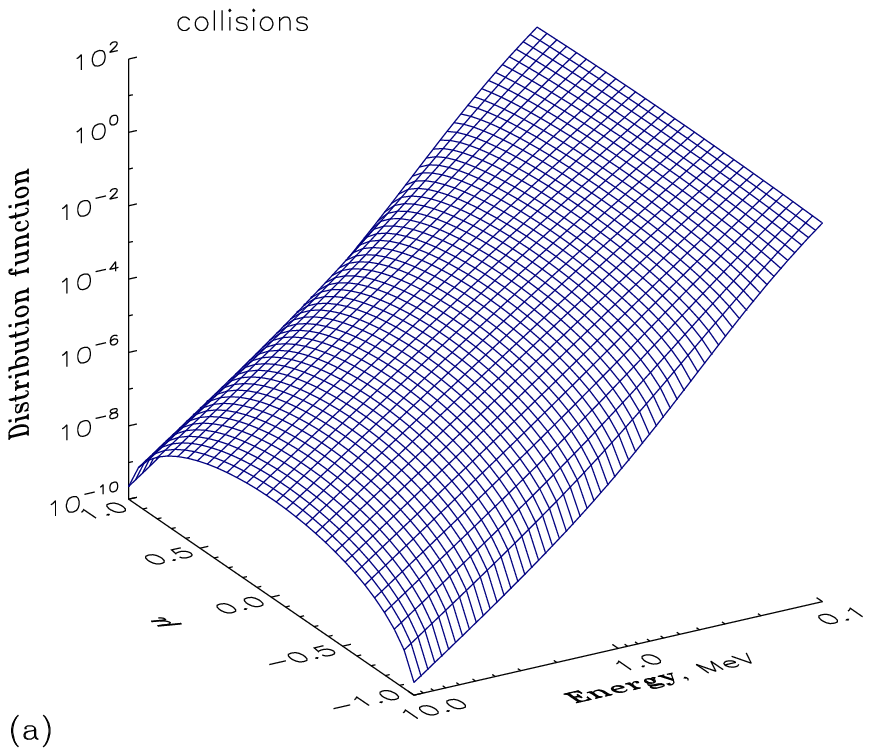}%
\includegraphics{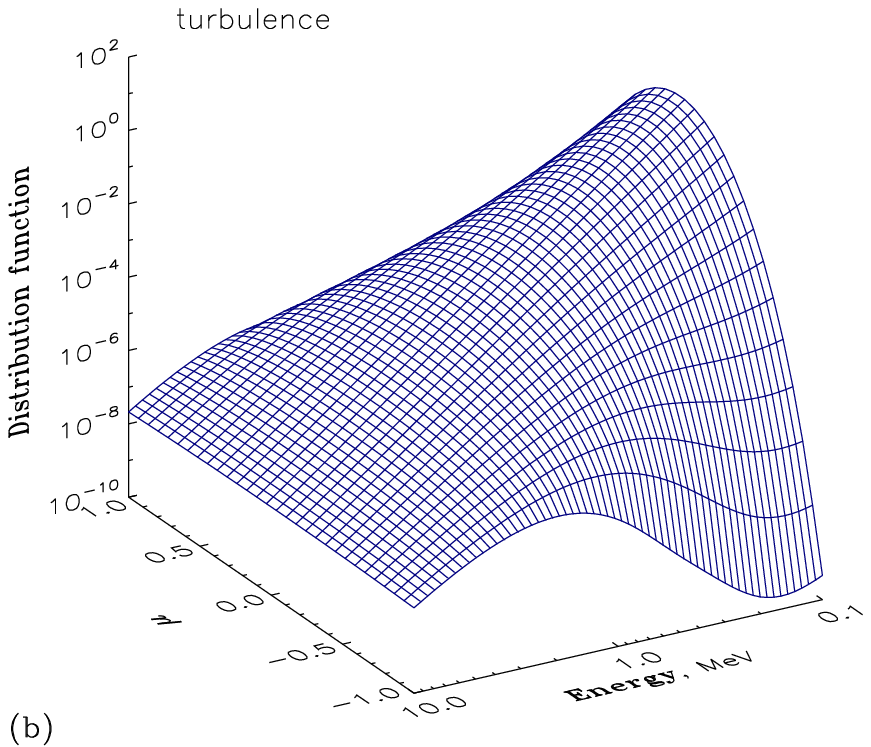}
\caption{Electron distribution functions affected by collisions (a) or magnetic turbulence (b).}
\label{FigDFs}
\end{figure*}

We have applied the pitch-angle scattering model (\ref{mu_scattering}) with the scattering time given by (\ref{tau_c}) and shown in Figure \ref{FigEvolutionTau}b to an electron distribution function that initially was described by the following (factorized) analytical expression:
\begin{equation}\label{df0}
f_0(E, \mu)\propto E^{-\delta}\exp\left(-\frac{\mu^2}{\Delta\mu^2}\right)
\end{equation}
in the energy range from $E_{\min}=0.1$ MeV to $E_{\max}=10$ MeV, with $\delta=4$ and $\Delta\mu=0.04$; this function represents a power-law distribution over energy and a symmetric loss-cone (or ``pancake'') distribution over pitch-angle. We have assumed the integration time to be $t=28.87$ s, which is the minimum value of the scattering time in the considered energy range, or $t=\tau(E_{\min}$). The resulting electron distribution function is shown in Figure \ref{FigDFs}a. Here, the electron distribution at the \textit{low-energy} cutoff is almost isotropic, while at higher energies it retains a considerable anisotropy (equivalent to a loss-cone distribution with $\Delta\mu=0.552$).

\begin{figure*}
\includegraphics{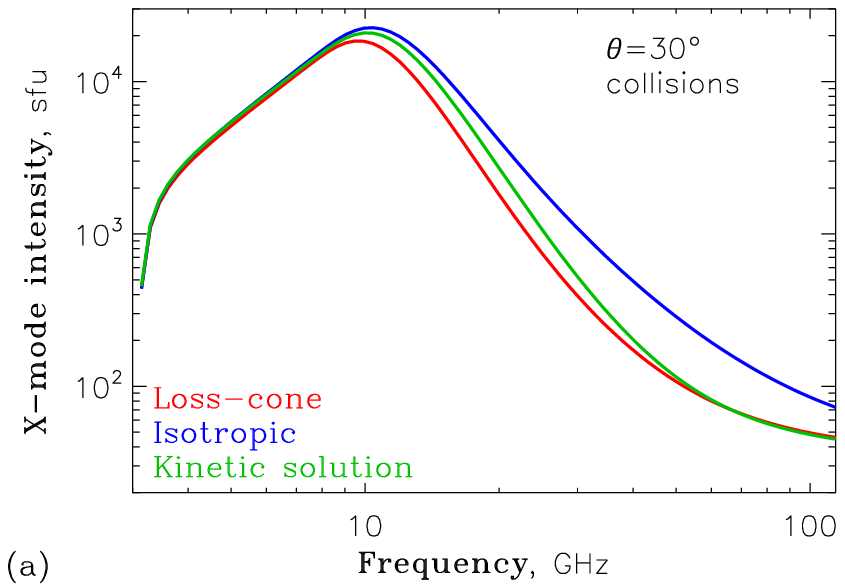}%
\includegraphics{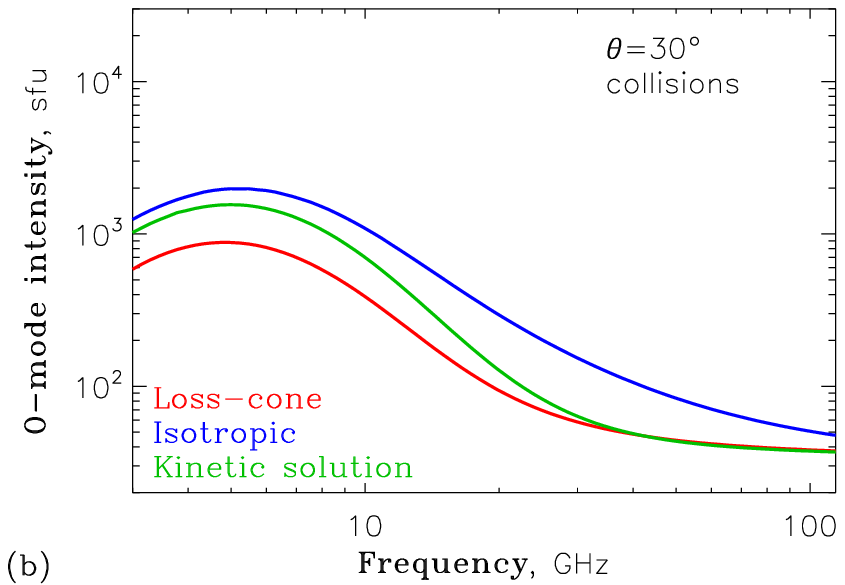}\\
\includegraphics{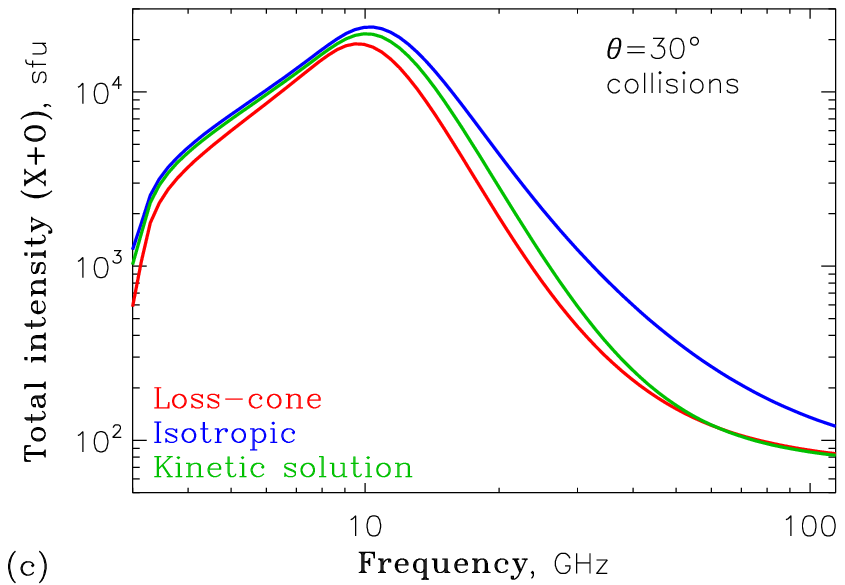}%
\includegraphics{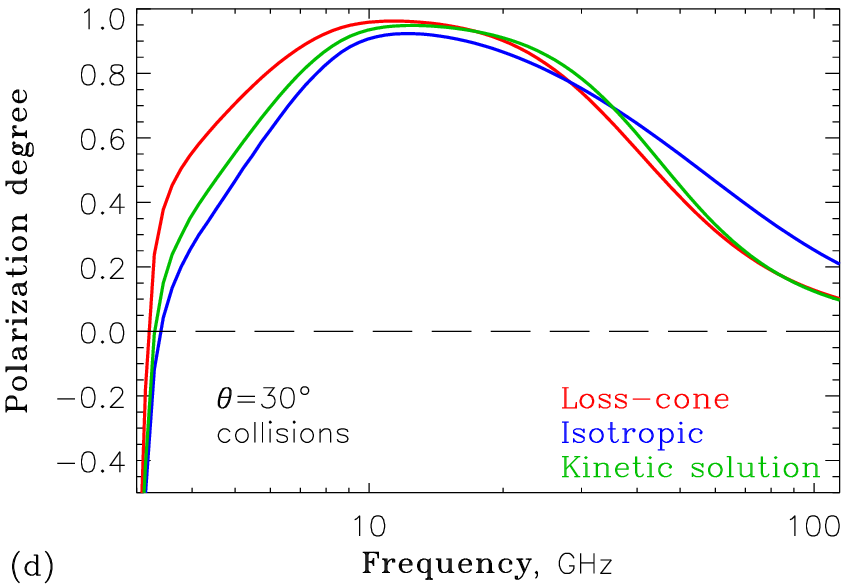}
\caption{Intensity and polarization spectra of microwave emission from an electron distribution with energy-dependent anisotropy (determined by collisions, see Figure \protect\ref{FigDFs}a); emission spectra for the isotropic distribution and loss-cone with $\Delta\mu=0.552$ (equivalent to the pitch-angle distributions at the low- and high-energy cutoffs) are shown, too. Simulation parameters: magnetic field $B=1000$ G, thermal electron density $n_0=10^{10}$ $\textrm{cm}^{-3}$, plasma temperature $T=10$ MK, nonthermal electron density $n_{\mathrm{b}}=10^3$ $\textrm{cm}^{-3}$, viewing angle $\theta=30^{\circ}$.}
\label{FigCollisionSpectra30}
\end{figure*}

\begin{figure*}
\includegraphics{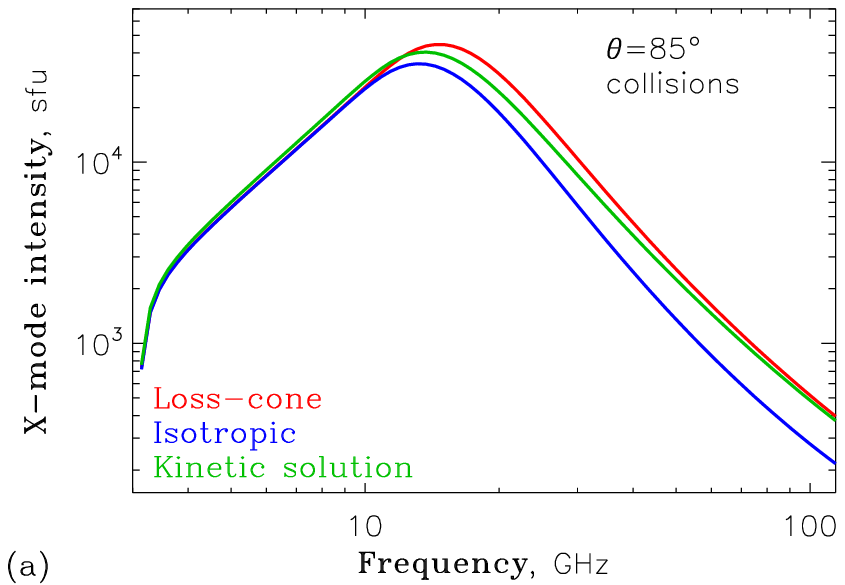}%
\includegraphics{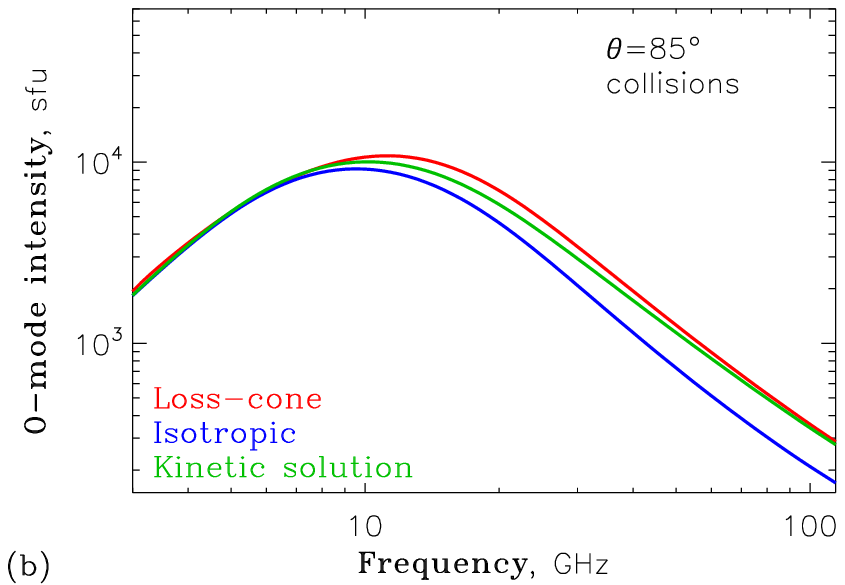}\\
\includegraphics{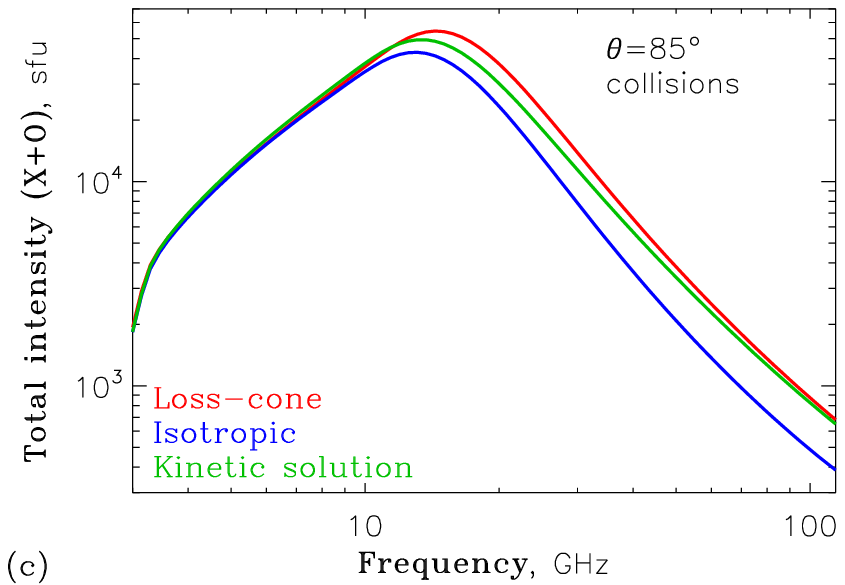}%
\includegraphics{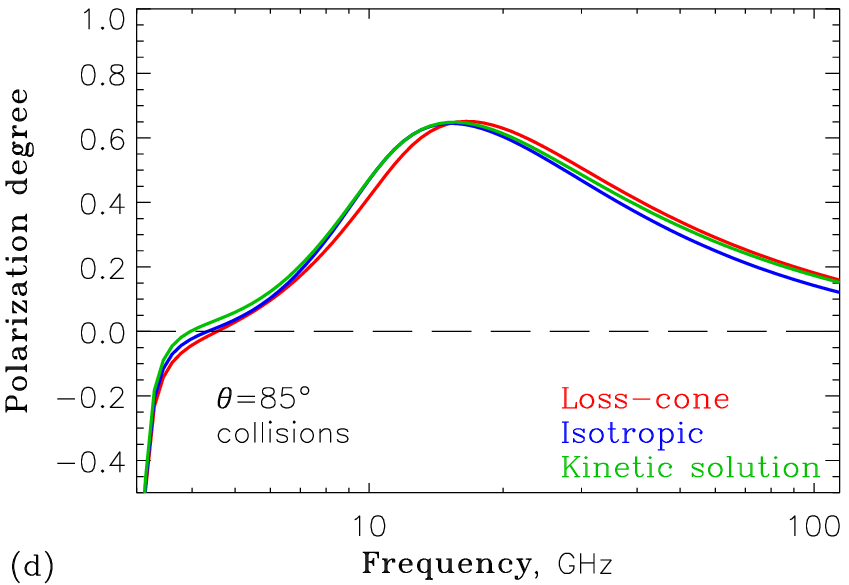}
\caption{Same as in Figure \protect\ref{FigCollisionSpectra30}, for the viewing angle of $\theta=85^{\circ}$.}
\label{FigCollisionSpectra85}
\end{figure*}

The corresponding microwave spectra for two different viewing angles are shown in Figures \ref{FigCollisionSpectra30}--\ref{FigCollisionSpectra85}; {the spectra were computed within the continuous approximation (the harmonic structure at low frequencies is ignored).} For comparison, we also show the simulation results for factorized electron distributions with the same pitch-angle profiles at all energies: the isotropic distribution and the loss-cone with $\Delta\mu=0.552$ (which are equivalent to the pitch-angle profiles of the non-factorized distribution function in Figure \ref{FigDFs}a at the low- and high-energy ends of the electron energy spectrum, respectively). One can see that the emission parameters for the non-factorized electron distribution are between those for the isotropic and loss-cone distributions; at higher frequencies, the emission parameters for the non-factorized electron distribution approach those for the loss-cone distribution. We note that the mentioned effect is noticeable in strong magnetic fields ($B\gtrsim 1000$ G) only; in weaker fields, the deviation from the loss-cone distribution is negligible because only the high-energy electrons make a contribution.

\subsection{Scattering on turbulence}
In the presence of magnetic fluctuations, the mean free path $\lambda_\mathrm{t}$ of energetic electrons 
is often approximated by a power-law dependence on energy \citep[e.g.,][]{2018A&A...610A...6M}:
\begin{equation}\label{lambdaT}
\lambda_{\mathrm{t}}\propto E^{-\alpha};
\end{equation}
accordingly, the {isotropization} time is given by
\begin{equation}\label{tauT}
\tau_{\mathrm{t}}=\frac{\lambda_{\mathrm{t}}}{v}.
\end{equation}
Therefore, the \textit{high-energy} electrons are scattered more efficiently. \citet{2018A&A...610A...6M} analyzed the X-ray and microwave source sizes in the 21 May 2004 flare (which were assumed to be determined by scattering of nonthermal electrons on turbulence) and estimated the corresponding scattering mean free path values as $\lambda\simeq 1400$ km at $E\simeq 25$ keV and $\lambda\simeq 100$ km at $E\simeq 400$ keV; this corresponds to a power-law index of $\alpha\simeq 0.95$ in Equation (\ref{lambdaT}) {if we associate the estimated mean free path $\lambda$ with the turbulent one $\lambda_\mathrm{t}$}. The energy dependence of the turbulent scattering time $\tau_\mathrm{t}$ is shown in Figure \ref{FigEvolutionTau}b.

We have applied the pitch-angle scattering model (\ref{tau_c}) with the scattering described by Equations (\ref{lambdaT}--\ref{tauT}) and estimations by \citet{2018A&A...610A...6M} (with the resulting scattering time shown in Figure \ref{FigEvolutionTau}b) to evolve the electron distribution function described initially by Equation (\ref{df0}) in the energy range from $E_{\min}=0.1$ MeV to $E_{\max}=10$ MeV, with $\delta=4$ and $\Delta\mu=0.2$ (note that the initial angular distribution is wider than in the previous example). We have adopted the integration time of $t=1.56\times 10^{-5}$ s, which is the minimum value of the scattering time in the considered energy range, or $t=\tau(E_{\max}$). The resulting electron distribution function is shown in Figure \ref{FigDFs}b. Here, the electron distribution at the \textit{high-energy} end is almost isotropic, while at lower energies it retains a considerable anisotropy (equivalent to a loss-cone distribution with $\Delta\mu=0.231$).

\begin{figure*}
\includegraphics{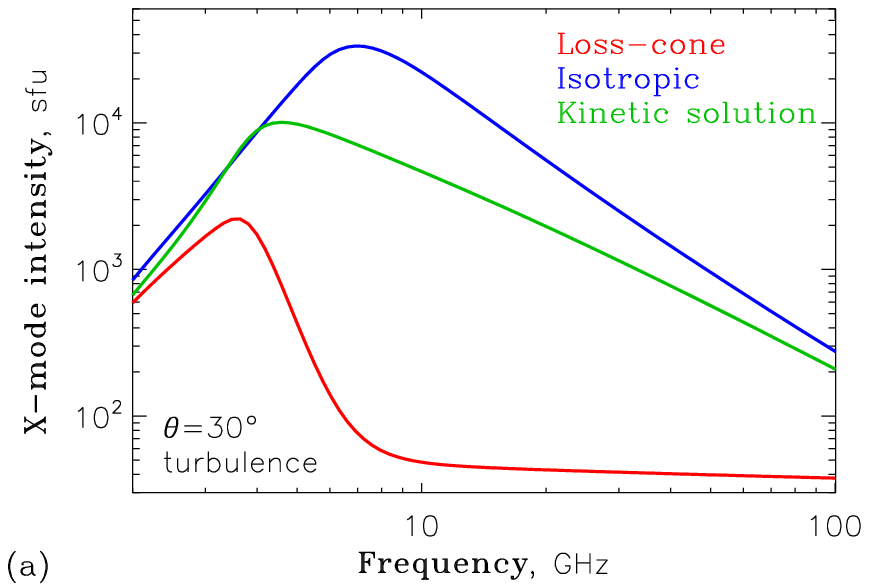}%
\includegraphics{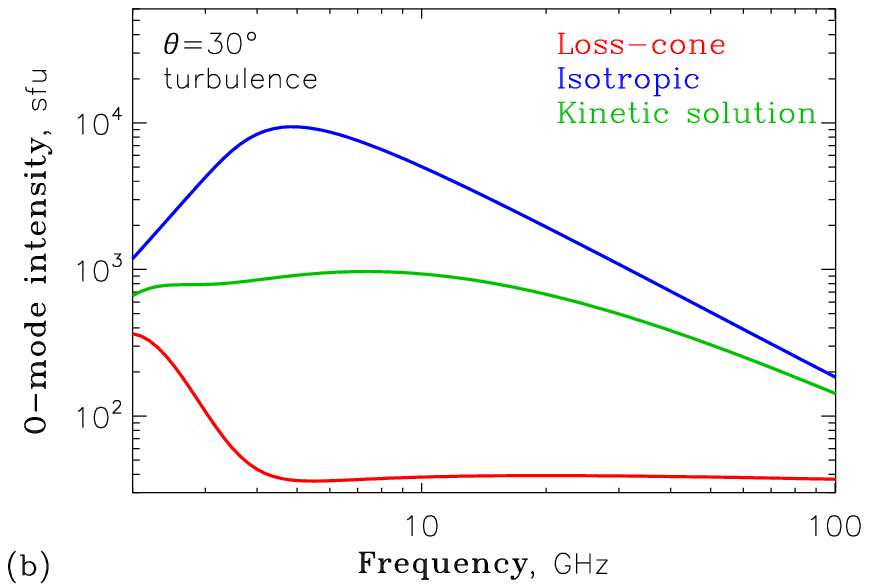}\\
\includegraphics{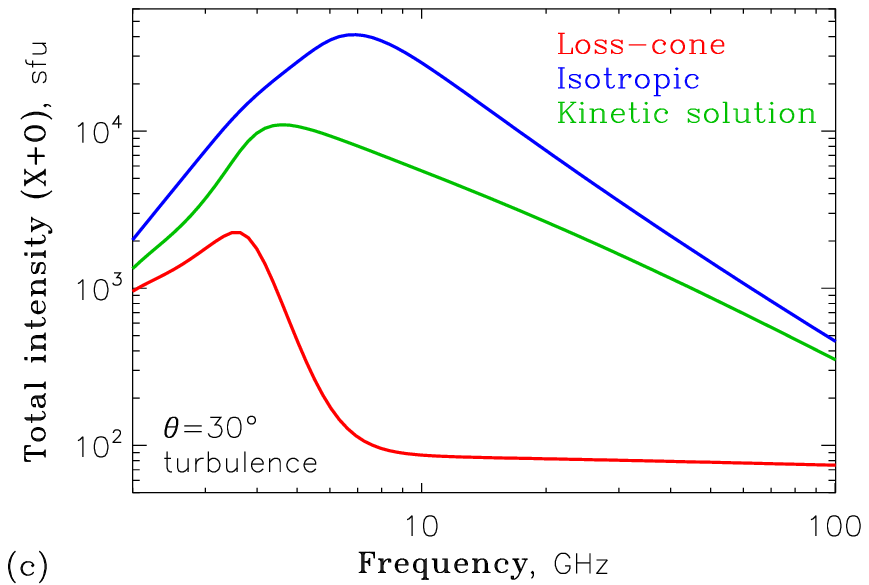}%
\includegraphics{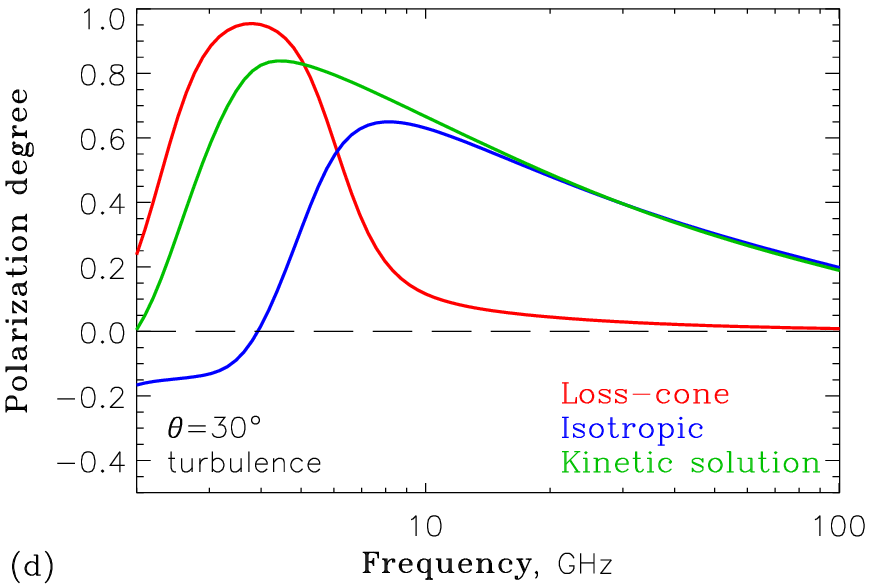}
\caption{Intensity and polarization spectra of microwave emission from an electron distribution with energy-dependent anisotropy (determined by turbulence, see Figure \protect\ref{FigDFs}b); emission spectra for the isotropic distribution and loss-cone with $\Delta\mu=0.231$ (equivalent to the pitch-angle distributions at the high- and low-energy cutoffs) are shown, too. Simulation parameters: magnetic field $B=180$ G, thermal electron density $n_0=10^{10}$ $\textrm{cm}^{-3}$, plasma temperature $T=10$ MK, nonthermal electron density $n_{\mathrm{b}}=10^6$ $\textrm{cm}^{-3}$, viewing angle $\theta=30^{\circ}$.}
\label{FigTurbulenceSpectra30}
\end{figure*}

\begin{figure*}
\includegraphics{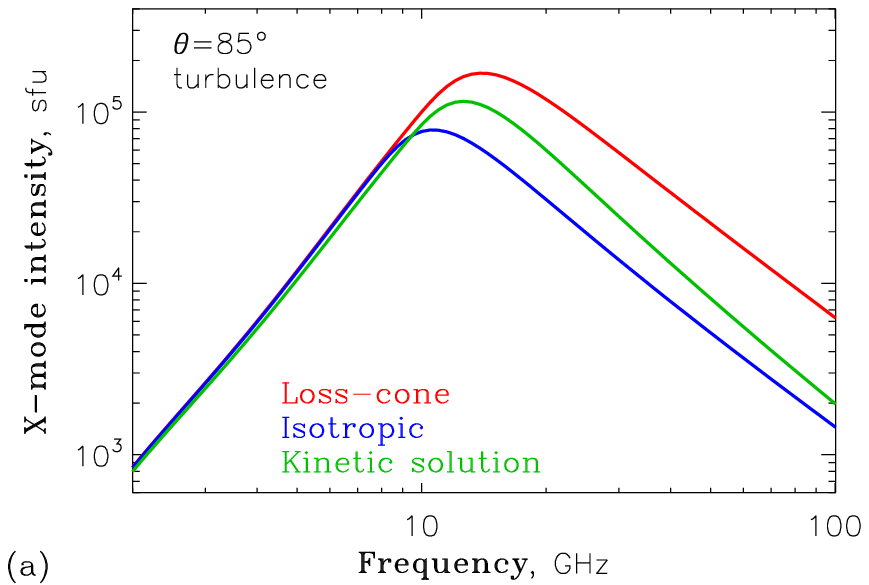}%
\includegraphics{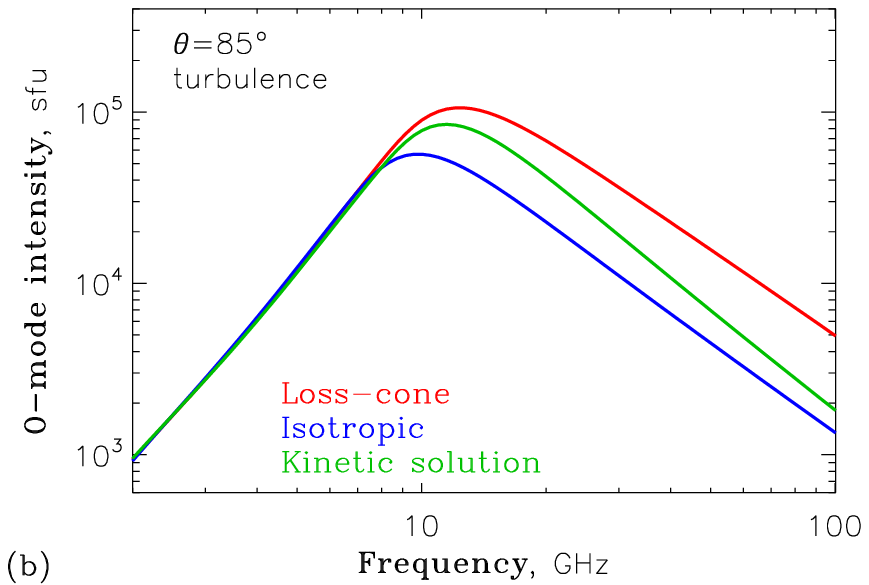}\\
\includegraphics{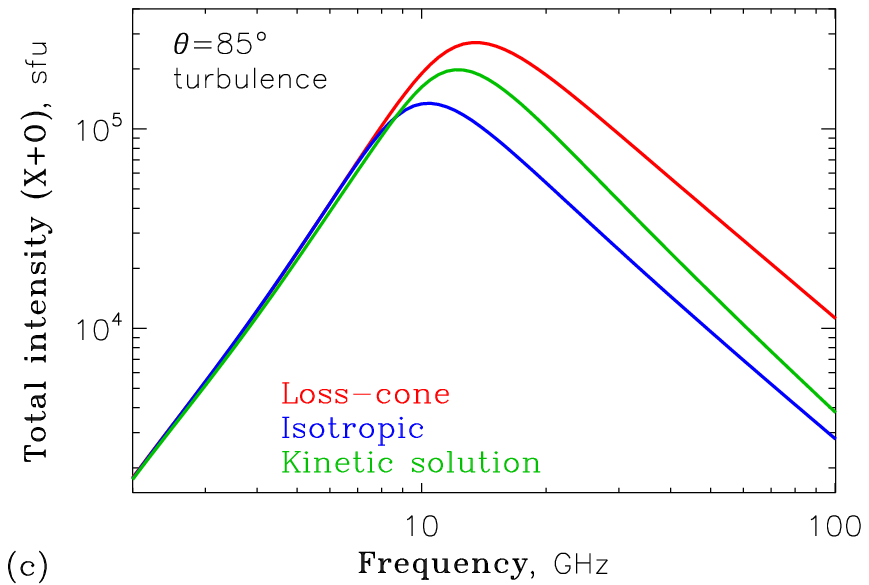}%
\includegraphics{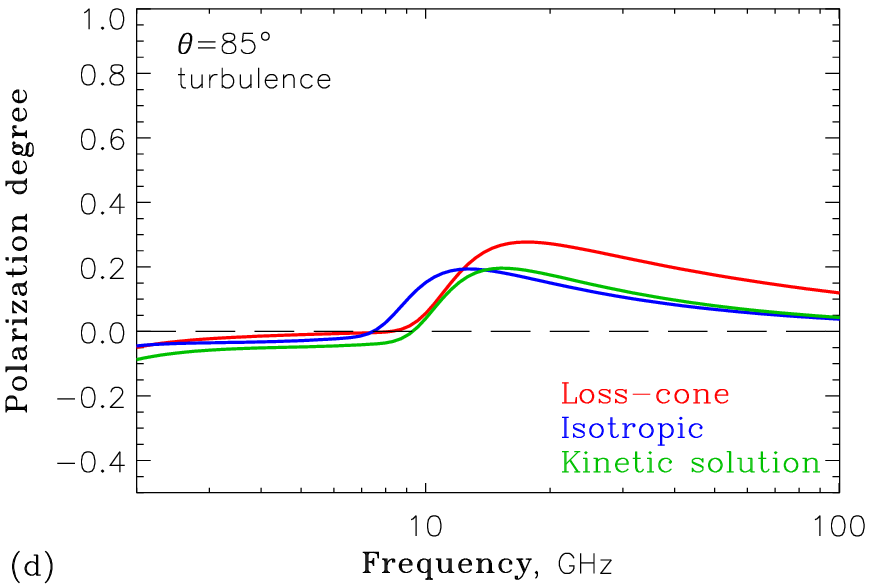}
\caption{Same as in Figure \protect\ref{FigTurbulenceSpectra30}, for the viewing angle of $\theta=85^{\circ}$.}
\label{FigTurbulenceSpectra85}
\end{figure*}

The corresponding microwave spectra, {again computed within the continuous approximation,} for two different viewing angles are shown in Figures \ref{FigTurbulenceSpectra30}--\ref{FigTurbulenceSpectra85}; for comparison, we also show the simulation results for factorized electron distributions with the same pitch-angle profiles at all energies: the isotropic distribution and the loss-cone with $\Delta\mu=0.231$ (which are equivalent to the pitch-angle profiles of the non-factorized distribution function in Figure \ref{FigDFs}b at the high- and low-energy ends, respectively). One can see that the emission parameters for the non-factorized electron distribution are again between those for the isotropic and loss-cone distributions; at higher frequencies, emission parameters for the non-factorized electron distribution approach those for the isotropic distribution. The mentioned effect is well visible for the typical coronal magnetic field strengths ($B\sim 100$ G). 

\subsection{Spectral indices}\label{Indices}
\begin{figure*}
\includegraphics{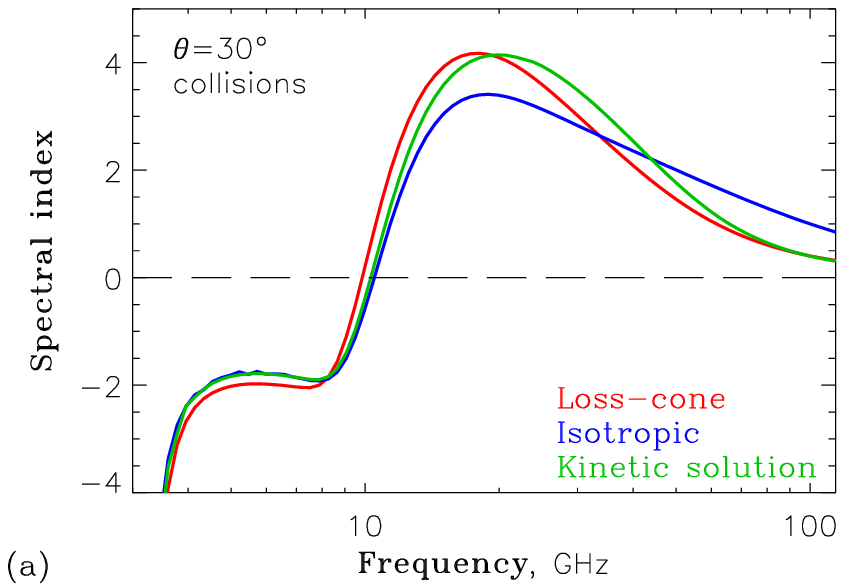}%
\includegraphics{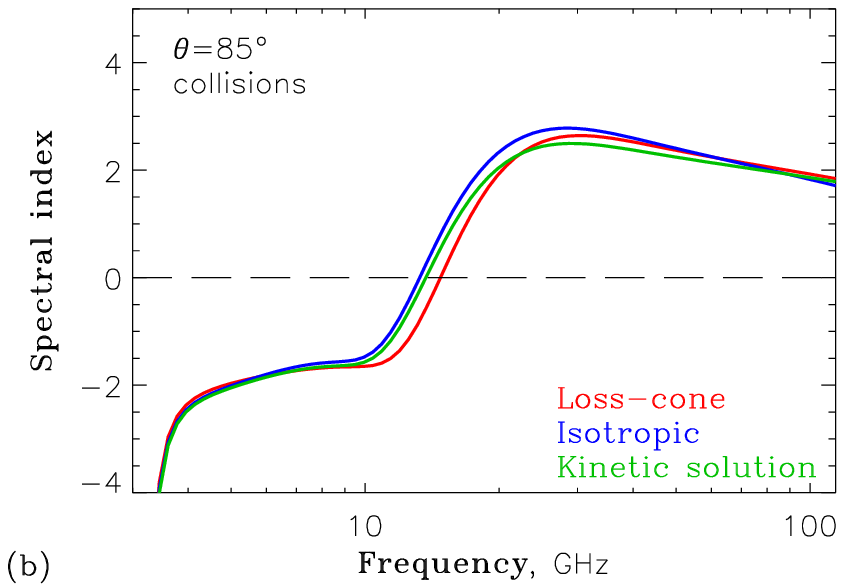}
\caption{Spectral indices of microwave emission computed for the same electron distributions (including the distribution with energy-dependent anisotropy determined by collisions) and source parameters as in Figures \protect\ref{FigCollisionSpectra30}--\protect\ref{FigCollisionSpectra85}; the viewing angles are $\theta=30^{\circ}$ (a) and $85^{\circ}$ (b).}
\label{indicesC}
\end{figure*}

\begin{figure*}
\includegraphics{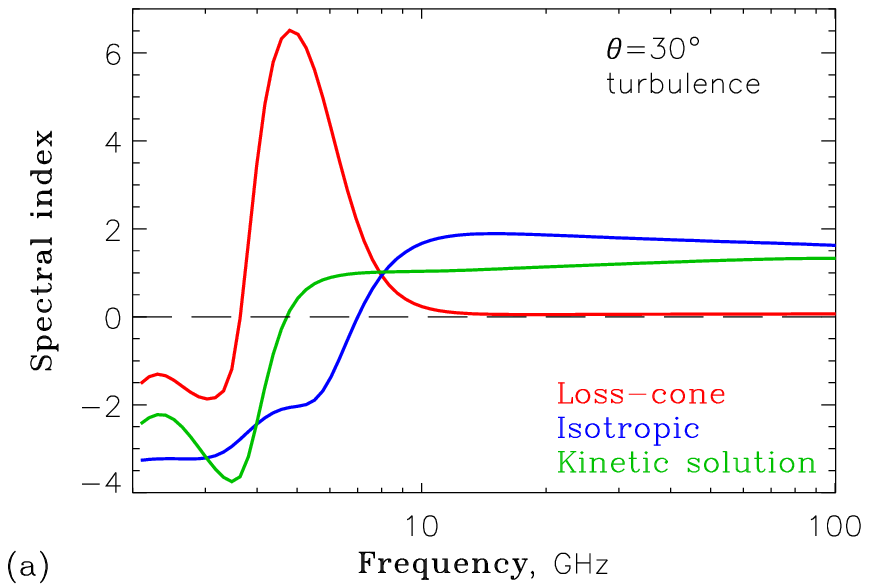}%
\includegraphics{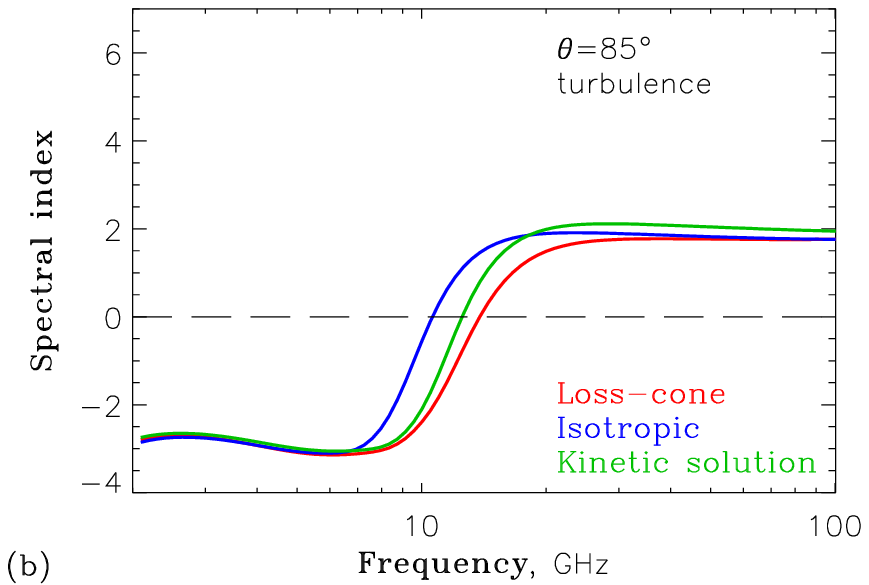}
\caption{Spectral indices of microwave emission computed for the same electron distributions (including the distribution with energy-dependent anisotropy determined by turbulence) and source parameters as in Figures \protect\ref{FigTurbulenceSpectra30}--\protect\ref{FigTurbulenceSpectra85}; the viewing angles are $\theta=30^{\circ}$ (a) and $85^{\circ}$ (b).}
\label{indicesT}
\end{figure*}

Figures \ref{indicesC} and \ref{indicesT} demonstrate the spectral indices of microwave emission computed using the above models. The spectral index is defined as
\begin{equation}
\delta_\mu=-\frac{f}{I}\frac{\partial I}{\partial f}.
\end{equation}
The electron distribution functions and other source parameters in Figures \ref{indicesC} and \ref{indicesT} are the same as in Figures \ref{FigCollisionSpectra30}--\ref{FigCollisionSpectra85} and \ref{FigTurbulenceSpectra30}--\ref{FigTurbulenceSpectra85}, respectively. In all cases, the spectral index computed from the kinetic solution differs measurably from those from the factorized approximations.

In the collision-dominated case (Figure \ref{indicesC}), the spectral indices for the non-factorized distribution at high frequencies are not much different from those for the loss-cone distribution, because the high-energy electrons (which are still anisotropic, despite of collisions) make a dominant contribution to the microwave emission. The largest deviation from the loss-cone case occurs at intermediate frequencies (slightly above the spectral peak). We also note that, with frequency, for small viewing angles the spectral index for the non-factorized distribution approaches that for the loss-cone distribution from above, while for large viewing angles the spectral index for the non-factorized distribution approaches that for the loss-cone distribution from below.

In contrast, in the turbulence-dominated case (Figure \ref{indicesT}), the spectral indices for the non-factorized distribution at high frequencies are more similar to those for the isotropic distribution. At small viewing angles, the optically thin spectral index for the non-factorized distribution is smaller (i.e., the spectrum is less steep) than that for the isotropic distribution, although this difference decreases with frequency. At large viewing angles, the differences between the considered electron distributions become very small but still noticeable: for the non-factorized distribution, the optically thin spectral index is larger (i.e., the spectrum is steeper) than those for the loss-cone and isotropic distributions. 

\section{Discussion and Conclusions}\label{sec_Disc}
In this paper we extend our fast GS codes \citep{Fl_Kuzn_2010} to the case, where the distribution of the nonthermal electrons can be defined numerically in the form of two-dimensional arrays, describing the dependence of the nonthermal electrons on the energy and pitch-angle. Along with this feature, we made several other improvements and additions. In particular, we improved treatment of the free-free component following the theory of \citet{2021ApJ...914...52F} and permitted arbitrary user-defined list of frequencies where to compute the microwave emission.

The main development is the ability of the new codes to deal with the numerically defined distributions of the nonthermal electrons. This development paves the way for direct use of numeric solutions of acceleration/transport equations/codes to compute the observables such as the flux and polarization of the microwave emission.

Such solutions are now routinely obtained in numerical simulations of various levels of complexity---from `toy' models, e.g. within a simple `escape time' approximation \citep{Fl_2005n} to sophisticated models, such as \textit{kglobal} \citep{PhysRevLett.126.135101}, that take many important physical effects into account and permit arbitrary nonfactorized anisotropic distribution of the nonthermal electrons to be considered. The new codes offer the computation speed comparable to that of \citet{Fl_Kuzn_2010} supporting comparably fast computation of the microwave emission from rather large models.

\acknowledgments  This work was supported in part by NSF grants AGS-1817277 and 
AGS-2121632, 
and NASA grants
80NSSC18K0667, 
80NSSC19K0068, 
80NSSC18K1128 
and 80NSSC20K0627 
to New Jersey Institute of Technology, and by the Ministry of Science and Higher Education of the Russian Federation. The authors thank Dr. Chen Bin for help with the code implementation for Mac OS and Python.

\bibliographystyle{apj}
\bibliography{demon,fleishman,gary_2018,natu_lett,magnetography,Kuznetsov,ugsc}

\appendix
\section{On the interpolation}\label{Interpolation}
The numerically-defined electron distribution function is specified by a 2D array of values $f_{ij}=f(E_i, \mu_j)$ at the nodes of a regular grid given by $E_0, E_1, \ldots, E_{M-1}$ and $\mu_0, \mu_1, \ldots, \mu_{N-1}$. To compute the gyrosynchrotron emission, we need to know the function values $f$, together with its derivatives over energy $f'_E$ and pitch-angle $f'_{\mu}$ and (for the continuous gyrosynchrotron approximation) the second derivative over pitch-angle $f''_{\mu\mu}$, at arbitrary points $(E, \mu)$ in the energy-pitch-angle space. For this, we use two different interpolation methods.

\paragraph{2D cubic spline interpolation}
The classical implementation of 1D cubic spline interpolation \citep[see, e.g.,][]{MathRec} implies that at any interval $x_i\le x<x_{i+1}$, the interpolated function $f(x)$ is given by a local cubic polynomial depending on the specified function values at the adjacent grid nodes $f(x_i)$ and $f(x_{i+1})$ and on the second derivatives of the function at the same nodes $f''(x_i)$ and $f''(x_{i+1})$; the second derivatives at all nodes are computed globally (before the interpolation) using the specified function values at the nodes, by solving a system of equations. Accordingly, the first and second derivatives of the interpolated function $f'(x)$ and $f''(x)$ are given by local quadratic and linear polynomials, respectively. In our code, the 2D spline interpolation is implemented in the following way: at the initialization step, before the interpolation, we construct a number of 1D splines, i.e.
\begin{enumerate}
\item
For each column of the input array (i.e., for $E=E_i$), we construct a spline describing the pitch-angle dependence of the distribution function $f=f(E_i, \mu)$, and compute the second derivative values $f''_{\mu\mu}(E_i, \mu_j)$ at the grid nodes.
\item
For each row of the input array (i.e., for $\mu=\mu_j$), we construct a spline describing the energy dependence of the distribution function $f=f(E, \mu_j)$, and compute the second derivative values $f''_{EE}(E_i, \mu_j)$ at the grid nodes.
\item
Using the results of the previous step, for each column of the input array, we construct a spline describing the pitch-angle dependence of the second derivative of the distribution function over energy $f''_{EE}=f''_{EE}(E_i, \mu)$ and compute the fourth derivative values $f''''_{EE\mu\mu}(E_i, \mu_j)$ at the grid nodes.
\end{enumerate}
At the boundaries of the array, the first derivatives of the 1D spline functions are assumed to be equal those derived using three-point Lagrange interpolation---this approach was found to provide the most accurate results.

At each point $(E, \mu)$, such that $E_i\le E<E_{i+1}$ and $\mu_j\le\mu<\mu_{j+1}$, the interpolation is performed in two steps. Firstly, we interpolate over pitch-angle $\mu$, i.e., compute the values of the distribution function and its derivatives at the points $(E_i, \mu)$ and $(E_{i+1}, \mu)$:
\begin{enumerate}
\item
Using the values $f(E_i, \mu_j)$, $f(E_i, \mu_{j+1})$, $f''_{\mu\mu}(E_i, \mu_j)$, and $f''_{\mu\mu}(E_i, \mu_{j+1})$, and the 1D spline formulae, we compute the interpolated values of the distribution function $f(E_i, \mu)$, as well as of its derivatives over pitch-angle $f'_{\mu}(E_i, \mu)$ and $f''_{\mu\mu}(E_i, \mu)$.
\item
Using the values $f''_{EE}(E_i, \mu_j)$, $f''_{EE}(E_i, \mu_{j+1})$, $f''''_{EE\mu\mu}(E_i, \mu_j)$, and $f''''_{EE\mu\mu}(E_i, \mu_{j+1})$, and the 1D spline formulae, we compute the interpolated values of the second derivative of the distribution function over energy $f''_{EE}(E_i, \mu)$, as well as of its derivatives over pitch-angle $f'''_{EE\mu}(E_i, \mu)$ and $f''''_{EE\mu\mu}(E_i, \mu)$.
\end{enumerate}
Similarly, we compute the corresponding interpolated values at $E=E_{i+1}$: $f(E_{i+1}, \mu)$, $f'_{\mu}(E_{i+1}, \mu)$, $f''_{\mu\mu}(E_{i+1}, \mu)$, $f''_{EE}(E_{i+1}, \mu)$, $f'''_{EE\mu}(E_{i+1}, \mu)$, and $f''''_{EE\mu\mu}(E_{i+1}, \mu)$. Then we perform interpolation over energy $E$, i.e.
\begin{enumerate}
\item
Using the values $f(E_i, \mu)$, $f(E_{i+1}, \mu)$, $f''_{EE}(E_i, \mu)$, and $f''_{EE}(E_{i+1}, \mu)$, and the 1D spline formulae, we compute the interpolated values of the distribution function $f(E, \mu)$ and of its derivative $f'_E(E, \mu)$.
\item
Using the values $f'_{\mu}(E_i, \mu)$, $f'_{\mu}(E_{i+1}, \mu)$, $f'''_{EE\mu}(E_i, \mu)$, and $f'''_{EE\mu}(E_{i+1}, \mu)$, and the 1D spline formula, we compute the interpolated values of the first derivative of distribution function over pitch-angle $f'_{\mu}(E, \mu)$.
\item
Using the values $f''_{\mu\mu}(E_i, \mu)$, $f''_{\mu\mu}(E_{i+1}, \mu)$, $f''''_{EE\mu\mu}(E_i, \mu)$, and $f''''_{EE\mu\mu}(E_{i+1}, \mu)$, and the 1D spline formula, we compute the interpolated values of the second derivative of distribution function over pitch-angle $f''_{\mu\mu}(E, \mu)$.
\end{enumerate}

\paragraph{Linear-quadratic 2D interpolation}
In contrast to the previous case, this is a local approach, when the interpolated function values are determined by its values at several (four or six) adjacent nodes. The distribution function $f(E, \mu)$ itself is computed using bilinear interpolation. The derivatives $f'_E(E, \mu)$ and $f'_{\mu}(E, \mu)$ are computed using a linear interpolation in one variable and a Lagrange interpolation on three closest points in another variable (where the derivative is needed).

By default, we use the 2D cubic spline interpolation, because it usually provides a higher accuracy and speed (due to a higher smoothness of the interpolated function, the numerical routines converge faster). However, due to its global nature, this approach can produce undesirable artifacts (``wiggles'' on the interpolated function) when applied to distributions with very sharp gradients; in particular, a maser instability can occur even if the numerically-defined distribution function has no positive slope (but has a very steep negative slope somewhere). In this respect, the linear-quadratic interpolation is more stable. 

Another important feature related to interpolation is the grid spacing. Since the electron distributions in the solar corona can cover a wide range of energies,  logarithmically-spaced grids ($E_{i+1}/E_i=\textrm{const}$) are usually used to describe them. To improve the interpolation accuracy, in such cases we interpolate not the dependence of $f$ vs. $E$, but the dependence of $\log f$ vs. $\log E$; for a power-law distribution, both the spline and linear-quadratic interpolation methods provide exact results in this case. A disadvantage of this approach is that the distribution function array must not contain zero values. We offer a user the opportunity to switch between the $\log f$ vs. $\log E$ (suitable for logarithmically-spaced grids, used by default) and $f$ vs. $E$ (suitable for equidistant grids) interpolation regimes, according to their model. For the pitch-angle distributions, no variable substitutions are used, i.e., an equidistant grid over $\mu$ would provide the best results. We note that the values of $\mu$ should cover the entire range of possible values (i.e., $\mu_0=-1$ and $\mu_{N-1}=1$); otherwise, extrapolation beyond the specified range can occur during computations, which will reduce the accuracy greatly.
\end{document}